\documentclass[%
 reprint,
 amsmath,amssymb,
 aps,
]{revtex4-1}

\usepackage{graphicx}
\usepackage{dcolumn}
\usepackage{bm}
\usepackage{eqnarray,amsmath} 
\usepackage{xcolor}

\begin{document}

\preprint{APS/123-QED}

\title{Non-local effects reflect the jamming criticality in granular flows of frictionless particles}
\author{Hugo Perrin$^{1,2}$}
\author{Matthieu Wyart$^{2}$}%
\author{Bloen Metzger$^{1}$}%
\author{Yo\"el Forterre$^{1}$}%

\affiliation{$^1$Aix Marseille Univ., CNRS, IUSTI, 13453 Marseille, France \\
$^2$Institute of Physics, Ecole Polytechnique F\'ed\'erale de Lausanne, CH-1015 Lausanne, Switzerland}

\date{\today}

\begin{abstract}
The jamming transition is accompanied by a rich phenomenology, such as hysteresis or non-local effects, which is still not well understood.  Here we experimentally investigate a model frictionless granular layer flowing down an inclined plane, as a way to disentangle generic collective effects from those arising from frictional interactions. We find that thin frictionless granular layers are devoid of hysteresis, yet the layer stability is increased as it gets thinner. Rheological laws obtained for different layer thicknesses can be collapsed into a unique master curve, supporting that non-local effects are the consequence of the usual finite-size effects associated to the presence of a critical point. This collapse indicates that the so-called isostatic length $l^*$ governs the effect of boundaries on flow, and rules out other propositions made in the past. 

\end{abstract}

\maketitle

Dense amorphous media close to the solid/liquid transition present a rich phenomenology such as hysteresis, finite-size or non-local effects and dilatancy.  Understanding these phenomena,  which shape the jamming transition \cite{PhysRevE68011306,PhysRevLett99178001}, is a major challenge  to describe the flow of granular media \cite{forterre2008}, foams \cite{lespiat2011}, or soft glassy materials \cite{goyon2010} and is related to important phenomena in geophysics like earthquakes \cite{leeman2016}, landslides or avalanches \cite{lucas2014,ferdowsi2018}.  However the origin of these phenomena is still poorly understood and much debated. A striking example illustrating this situation is the well known fact that the critical angle $\theta_c$ at which a granular layer flows depends on its thickness $h$  \cite{Pouliquen1999,daerr1999two}. The diversity of interpretations and models put forward to explain this property, successively attributed to dilatancy effects \cite{Pouliquen1996}, to a diverging correlation length \cite{ertacs2002,pouliquen2004velocity, mills2008, gueudre2017scaling}, to the consequence of non-local rheology  
\cite{pouliquen2001slow,lemaitre2002origin,aranson2002continuum,pouliquen2009non,kamrin2015}, to boundary effects on mechanical stability \cite{wyart2009} or to hysteresis \cite{edwards2019}, highlights the difficulty in addressing such problem where many effects are potentially entangled.

Recently, we have developed a model granular system in which it is  experimentally possible to tune, and even eliminate, solid-friction between grains. This experimental control of the interparticle friction provided the mean to highlight the frictional transition at the origin of shear-thickening in dense particulate suspensions \cite{Clavaud2017}. More importantly, when investigated in a rotating drum configuration, this model granular system allowed us to show that both dilatancy effects and hysteresis of the avalanche angle disappear in the absence of inter-particle friction \cite{Clavaud2017,perrin2019}, confirming previous numerical predictions \cite{PeyneauRoux2008,degiuli2015}. Such a system of frictionless spheres thus provides a unique opportunity to study the flowing properties of granular layers without potential interplay with hysteretic behaviors or dilatancy effects.

About two decades ago, such an ideal granular material has been studied numerically by Peynaud and Roux \cite{PeyneauRoux2008}. Since then, the investigation of frictionless particulate systems in numerical simulations, whether in the inertial or viscous regimes, has brought major contributions to the theoretical understanding of granular and suspension flows \cite{combe2000,lerner2012,gallier2014,mari2014,wyart2014discontinuous,degiuli2015}. From a fondamental standpoint, this model granular material bridges frictional granular flows to other amorphous frictionless systems such as foams, emulsions and glassy materials. It thus provides the appealing possibility to discriminate features which are specific to frictional interactions, from those more generic that emanate from collective mechanical  effects.

In this paper, we investigate our model system of frictionless spheres in the inclined plane configuration, which enables us to both control the system size and impose a homogeneous friction coefficient $\mu$ to the medium. Our key findings are that (i)  hysteresis of the avalanche angle  is absent for frictionless grains whatever the thickness of the granular layer, showing that hysteresis solely arises from inter-particle friction and not from finite-size effects. By contrast, the layer stability  is strongly affected by finite-size effects. For frictionless grains, there is thus a unique critical curve  $h(\theta_c)$ separating flow from arrest. (ii) Flow rules for different layer thicknesses  collapse into a single master curve with a proper rescaling of variables. We explain this result from general finite size scaling arguments near a critical point \cite{cardy2012finite}, supporting that this language is appropriate to describe non-local effects in granular flows. (iii) This analysis supports that the length scale on which flow is affected by a boundary is the so-called isostatic length $\ell^*$ associated with the jamming transition \cite{wyart2005geometric}. Indeed, we argue that $h(\theta_c)\sim \ell^*\sim d\ (\tan \theta_c-\tan \theta_c^\infty)^{-\alpha}$, where  $\theta_c^\infty$ is the  threshold value of inclination for an infinite  thickness, with a predicted exponent $\alpha=0.83$ in good agreement with the experimental measurements  $\alpha\approx 0.9\pm 0.1$.

\paragraph*{{\bf Experimental set-up \& Protocols:}}
\begin{figure*}
\begin{center}
\includegraphics[width=16cm]{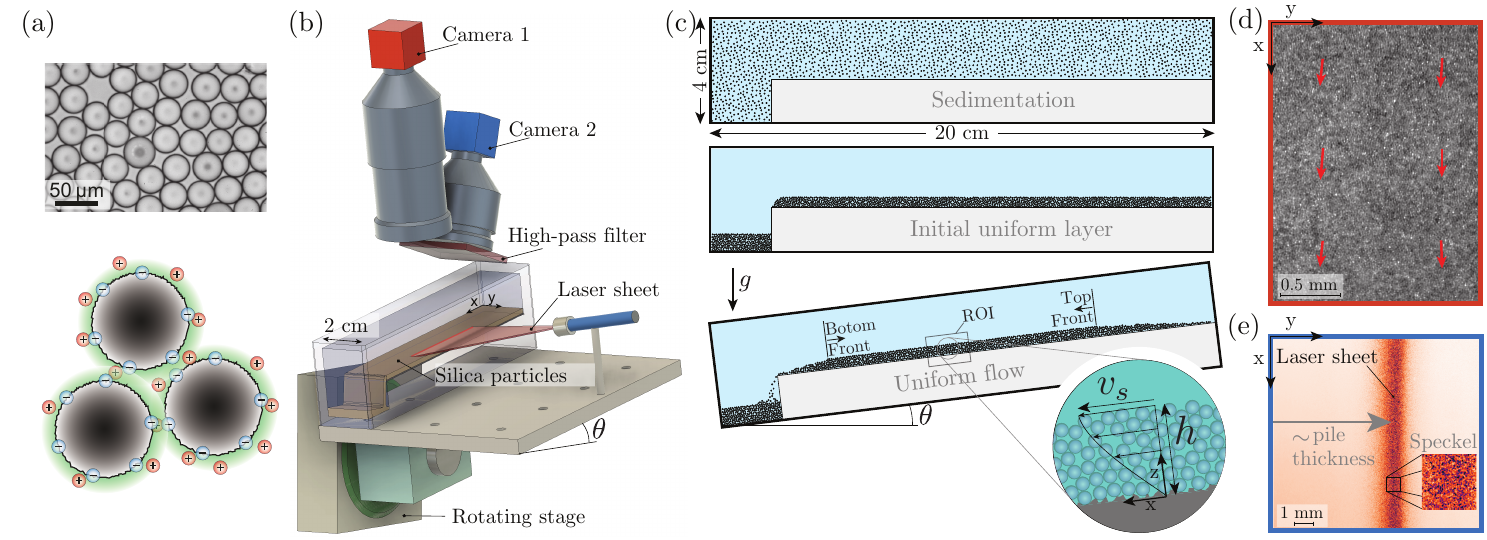}
\caption{Experimental set-up \& Protocols. (a) Top: Picture of the silica particles. Bottom: schematic of the electrostatic repulsive forces preventing frictional particle contacts. (b) Schematic of the experimental set-up. (c) Successive steps to prepare a uniform and pre-sheared layer of grains. (d) Image from camera 1 giving access to the particle surface velocity $v_s$ using PIV. (e) Image from camera 2 used to measure the pile thickness $h$ from the laser sheet position; and the pile stability from the time correlation of the laser speckle. \vspace{-0.7cm}
\label{fig1}}
\end{center}
\end{figure*}

We use a model frictionless granular system composed of silica spheres of diameter $d=23.46 \pm 1.05\;\rm{\mu m}$ [Fig. \ref{fig1}(a)]. When immersed in pure water, the surface of these particles is negatively charged. Under the low external stress used in this study (below 10 Pa), the resulting repulsive force \cite{Israelachvili2011Book} prevents the particles from making solid-frictional contacts. The particles then behave as frictionless spheres \cite{Clavaud2017,perrin2019}. To obtain such a behavior, the suspension must be cleaned in a piranha solution ($\rm{H_2O_2/H_2SO_4}=1/2$ by volume) for 10 minutes and then rinsed thoroughly with pure water. 

The set-up is shown in Fig. \ref{fig1}(b). The inclined plane is the bottom of a parallelepipedic container of size $20\times4\times2$ cm$^3$. It is covered with sand paper (Grade P500) to insure rough boundary conditions and avoid slippage. Both the inclined-plane and the cameras are mounted on a high precision rotation stage (PI M-060), which controls the inclination angle $\theta$. Prior to each experiment, the layer of grains is prepared as follows [see Fig. \ref{fig1}(c)]: after being resuspended, the particles sediment to form a uniform layer of thickness $\approx 30\,d$ on the incline. The plane is subsequently tilted at a large angle to pre-shear the suspension until the thickness of the layer reaches the desired value $h$. The inclinaison angle is then rapidly set-back to zero to stop the flow. As particles flow, two fronts form at the bottom and at the top of the incline, see Fig. \ref{fig1}(b). All measurements are performed in the region of interest (ROI) which is far from these two fronts and  where the flow is uniform.

To determine the rheology of the flowing layer of grains, we use the fact that for steady uniform flows down inclines  the suspension friction coefficient  is homogeneous across the layer thickness and given by the relation $\mu=\tau/P=\tan \theta$, where $\tau=\Delta \rho \phi g \sin\theta (h-z)$ is the shear stress, $P=\Delta \rho \phi g \cos\theta (h-z)$ is the normal stress, $\Delta \rho= 850$ kg$/$m$^{3}$ is the density difference between the particles and the suspending fluid, $\phi$  is the layer particle volume fraction and $g$ is the gravity. Moreover, by imaging the layer surface, camera 1 gives access to the grains surface velocity $v_s$ using PIV [Fig. \ref{fig1}(d)], and camera 2 tracks the transverse position of the inclined laser sheet [Fig. \ref{fig1}(e)], thereby giving access to the layer thickness $h$, with an absolute resolution of $\pm 2.5\;\rm{\mu m}$. These two measurements give access to the viscous number $J =\eta \dot \gamma / P= 2 \eta v_s / h^2 \phi \Delta \rho g \cos  \theta$,  assuming a parabolic velocity profile \cite{cassar2005submarine},  where $\dot \gamma=2  v_s / h$ is the shear rate and $\eta$ is the fluid viscosity.   Here, measurements are performed in the limit of small viscous number $J \leqslant 10^{-3}$; the volume fraction is thus assumed to be constant and given by the maximum packing of frictionless spheres $\phi = 0.64$. 

To characterize the stability of the flow and potential hysteretic behaviors of the avalanche angle, we need a quantitative way to discriminate whether, for a given angle $\theta$, the layer of grains is flowing or at rest. This constitutes an experimental challenge as the overdamped dynamics of the grains can be very slow and involve transients lasting several hours. Since our PIV measurements can only resolve grains surface velocities of the order $0.1\;\rm{\mu m / s}$, we developed a more sensitive speckle correlation technique \cite{CrassousCorrPRE2008}. The coherent laser light, when shined onto the granular layer composed of transparent beads, scatters randomly and produces a speckle [Fig. \ref{fig1}(e)]. Tiny movements of the particles are sufficient to change the optical path of the light, which sensibly decorrelates the speckle.  From successive images of camera 2, we can compute $C$, the space-averaged  time correlation of the speckle, see details in \footnote{C is the spatial average of the correlation matrix $M =  \frac{\left<I(0) I(\Delta t)\right> - \left<I(0)\right>\left<I(\Delta t)\right>}  {\sqrt{\left<I(0)^2\right> -\left<I(0)\right>^2 } \sqrt{\left<I(\Delta t)^2\right> -\left<I(\Delta t)\right>^2 } } $, where $I(0)$ and $I(\Delta t)$ are two images separated by a time interval $\Delta t = 0.1\;\rm{s}$ and $\left<.\right>$ denotes the average correlation over boxes of $4\times4$ pixels, see \cite{CrassousCorrPRE2008} for details}. For a layer of grains at rest, the correlation is slightly below 1 due to the intrinsic numerical noise of the camera. Its value is $C \approx 0.98$ with a typical noise of $\pm 0.01$. Flow can therefore be detected as soon as $C\leqslant C_{th}=0.96$, which typically corresponds to an average surface velocity of 0.5 nm$/$s. 

\begin{figure}
\begin{center}
\includegraphics[width=8cm]{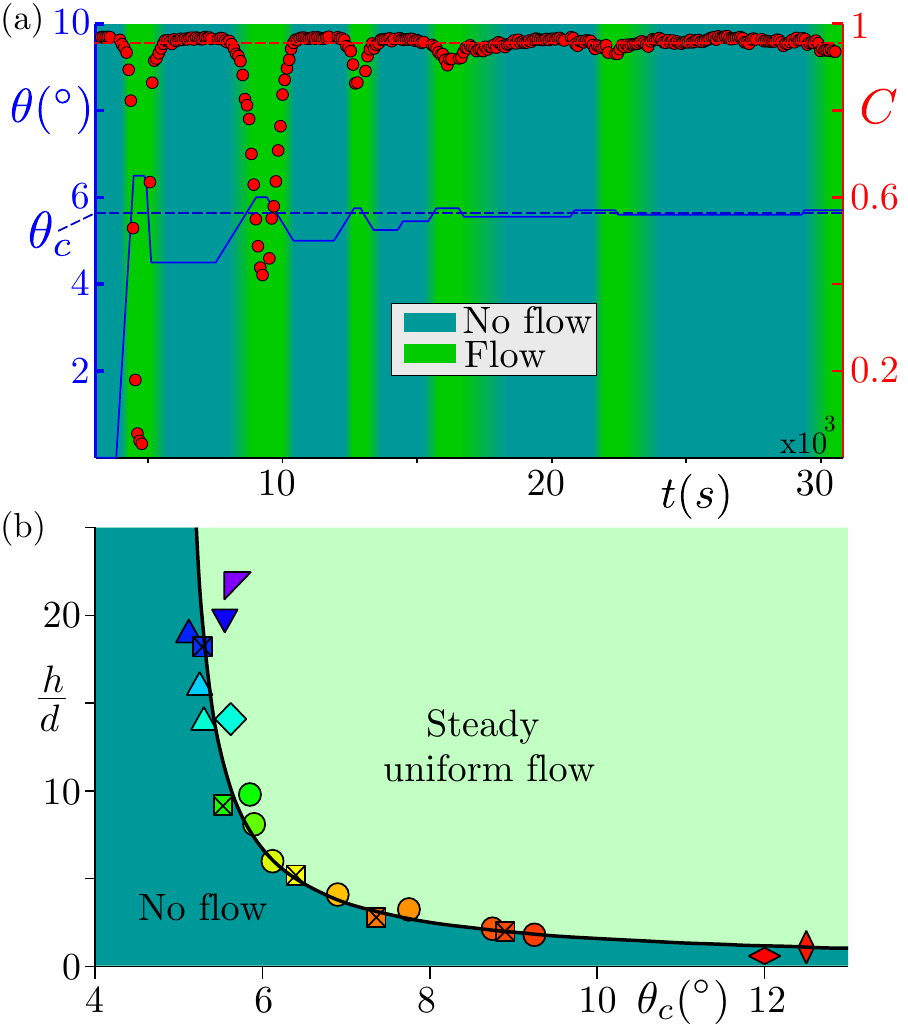}
\caption{Stability of a frictionless granular layer. (a) Dichotomy procedure: controlled inclination angle $\theta$ (in blue) and resulting speckle correlation $C$ (in red) versus time for a layer of thickness $h/d\simeq22$. Red-dashed line: $C_{\rm th}=0.96$ delimiting flow $(C<C_{\rm th})$ from arrest $(C>C_{\rm th})$. Blue-dashed line: critical stability angle $\theta_c(h/d)$. (b) Stability diagram: granular layer thickness $h/d$ versus critical stability angle $\theta_c$. 
Different open-markers indicate different runs, where the cleaning procedure and system aging induce an uncertainty on $\theta_c$ of the order of $0.5^\circ$. Crossed-markers correspond to the quasi-static friction coefficient $\mu_c(h/d)=\mu(J\to 0,h/d)$ deduced from the flow rules in Fig. \ref{fig4}(a). Black-line: best fit with $h/d = \left( (\tan \theta_c - \tan \theta_c^\infty )/a\right)^{-\alpha}$ yielding $\alpha = -0.9 \pm 0.1$, $a = 0.15$ and $\theta_c^\infty = 5.0^\circ$.\vspace{-0.7cm}}
\label{fig2}
\end{center}
\end{figure}

\noindent {\bf Influence of the system size on hysteresis and layer stability:} We first address whether finite-size frictionless systems exhibit hysteretic behavior of the flow onset. To this end, we developed the dichotomy procedure shown in Fig. \ref{fig2}(a).  Starting from a pre-sheared granular layer having the desired thickness $h$, the inclination angle $\theta$ is rapidly increased to $6.5^\circ$ and kept to a constant value. As shown in Fig. \ref{fig2}(a), the speckle correlation $C$ drops much below $C_{\rm th}$ (Red-dashed line) indicating that the granular layer starts flowing.  The angle of inclination $\theta$ is then successively decreased and increased in order to determine and gradually refine the frontier between flow and arrest. Importantly, the flow rates involved here are sufficiently small that during the whole dichotomy procedure, the layer thickness remains constant. As illustrated in Fig. \ref{fig2}(a), the dichotomy converges to a unique critical inclination angle $\theta_c(h/d)$  determined with a resolution of $\pm 0.05\rm{^\circ}$ (blue-dashed line),  and this for all initial layer thicknesses investigated [$1<h/d<22$]. These results show that a layer of frictionless particles stops or starts flowing at a unique critical angle $\theta_c(h/d)$~--~ the proof that no hysteresis of the avalanche angle is observed in frictionless granular systems, even of finite-size.

The critical stability angle $\theta_c(h/d)$ obtained with the above dichotomy procedure defined the stability diagram of the flow, which is plotted in Fig. \ref{fig2}(b). For granular layers of thickness larger than $10\,d$, the critical stability angle is nearly constant $ \theta_c^\infty \simeq 5^\circ $ and in good agreement with previous experimental ($6^\circ $ in \cite{perrin2019}) and numerical results ($5.73^\circ$ in \cite{PeyneauRoux2008}) obtained for frictionless systems in the infinite size limit. For thinner layers however, the critical stability angle $\theta_c$ increases and fitting the law  
\begin{equation}
h/d = \left( (\tan \theta_c - \tan \theta_c^\infty )/a\right)^{-\alpha},
\label{stabAng}
\end{equation}
yields $\alpha = 0.9 \pm 0.1$, $a = 0.15$ and $\theta_c^\infty = 5.0^\circ$. These results show that, conversely to hysteresis, finite-size effects are still in play and significantly impact the stability of frictionless granular layers.

\noindent{\bf Influence of the system size on rheology:} We now study how finite-size effects influence the granular layer dynamic flowing properties, in the aim of obtaining constitutive flow rules of the form $\mu(J,h/d)$. To this end, we fix the layer thickness $h$ and take advantage of the set-up to measure $J$ while imposing a quasi-static decreasing ramp of $\mu=\tan \theta$ using the rotation stage. This protocol provides, in a single measurement, the full rheological law $\mu(J,h/d)$ for a given value of $h/d$. We ensure that the decrease of the layer thickness during the ramp-down remains negligible (below $0.5\,d$) for the slow flows considered here ($J<10^{-3}$).  We also checked that steady state measurements provide the same results.

\begin{figure*}
\begin{center}
\includegraphics[width=14cm]{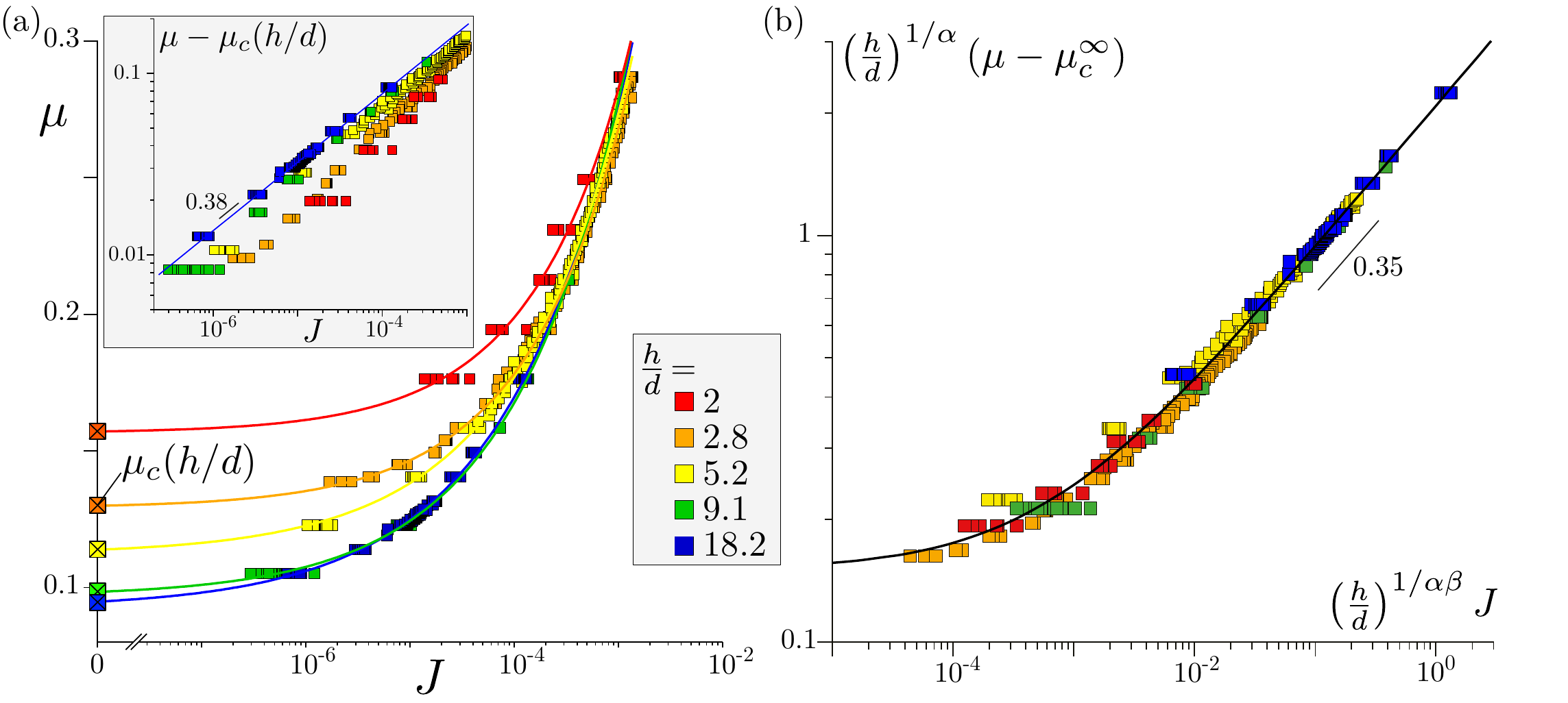}
\caption{Finite-size effects on the flow rules of a frictionless granular layer. (a) Macroscopic friction coefficient $\mu$ versus viscous number $J$ for different layer thicknesses $h/d$. Solid-lines: best fits with power laws plus constants. The critical friction coefficient $\mu_c(h/d)$ obtained when $J \to 0$ for each thickness (Crossed-squares) are reported in Fig. \ref{fig2}(b) using the relation $\mu_c(h/d)=\tan\theta_c(h/d)$. Inset: Same data plotting the reduced macroscopic friction $\mu-\mu_c(h/d)$ versus $J$. (b) Rescaled flow rule plotting $(h/d)^{1/\alpha}(\mu-\mu_c^\infty)$ versus $(h/d)^{1/\alpha\beta}J$ with $\alpha=0.9$ and $\beta=0.35$.   Black-line:  $g(x) = a + (c x ^{\gamma})( b x^{\beta})/(bx^{\beta}+ c x^{\gamma})$, with $a=0.15$, $b=2.0$, $c = 29$ and $\gamma = 0.73$.\vspace{-0.7cm}}
\label{fig4}
\end{center}
\end{figure*}

The rheological laws $\mu(J,h/d)$ are shown in Fig. \ref{fig4}(a) for various layer thickness $h/d$ and for a wide range of $J$.  We recover that the system size has a strong effect on the flow threshold. The quasi-static friction coefficient $\mu_c(h/d)=\mu(J\to 0,h/d)$ (crossed-markers) deduced from the flow rules is also fully consistent with the measurements of the critical stability angle obtained from the dichotomy procedure [see Fig. \ref{fig2}(b)], confirming the robustness of our finite-size effect characterization. We find that the system size not only affects the onset of the flow but also its dynamics, as evidenced by plotting the difference $\mu(J,h/d)-\mu_c(h/d)$ as function of $J$ [Fig. \ref{fig4}(a), inset].  
The large system ($h/d \simeq 18.2$) exhibits a power law with an exponent of $0.38$, in close agreement to the exponents measured experimentally in a rotating drum ($0.37 \pm 0.05$ \cite{perrin2019}) and derived theoretically ($\beta = 0.35$ \cite{degiuli2015}) for frictionless spheres in the infinite size limit. However, data for smaller systems increasingly deviate from this power law as the system size decreases. Together with Fig. \ref{fig2}(b) and Eq. \ref{stabAng}, these observations support the two following asymptotic regimes
\begin{eqnarray}
\mu\left(J \to 0 ,h/d\right)  &\sim&\mu_c^\infty + a (h/d)^{-1/\alpha}, \ \rm{and}\label{onset} \\
\mu\left(J ,h/d \to \infty \right) &\sim& \mu_c^\infty  + b J^{\beta}.\label{large}
\end{eqnarray}

To rationalise the observed behavior of $\mu(J,h/d)$, we use the framework of phase transitions  and the usual finite size scaling assumption valid near a critical point \cite{cardy2012finite}. 
In this view, $J$ is the order parameter controlled by the excess macroscopic friction $\Delta \mu \!\equiv \!\mu-\mu_c^\infty$ and system size $h$
\begin{equation}
\label{widom}
J(\Delta\mu,h)=\Delta\mu^{1/\beta} f(h/\xi(\Delta\mu)),
\end{equation}
where $f$ is some scaling function and $\xi(\Delta\mu)$ is a diverging length scale.  This scaling form is consistent with Eq. (\ref{onset}) if $\xi(\Delta\mu)=h(\theta_c)/c_0\sim d \Delta\mu^{-\alpha}$ where $c_0$ is some constant and $f(c_0)=0$, while Eq. (\ref{large}) imposes $f(\infty)=b^{-1/\beta}$. Eq. \ref{widom} can be rewritten as
\begin{equation}
\label{widom2}
\Delta\mu(J,h)=\left(h/d\right)^{-1/\alpha} g\left(\left(h/d\right)^{1/\alpha\beta} J\right),
\end{equation}
where $g$ is some function. This result implies that the data of Fig. \ref{fig4}(a) should collapse on a single curve by plotting the rescaled friction coefficient  $(h/d)^{1/\alpha} \Delta\mu(J,h)$ as function of the rescaled viscous number $\left(h/d\right)^{1/\alpha\beta} J$.  Remarkably, the collapse shown in Fig. \ref{fig4}(b) is excellent, without any fitting parameters [since $\beta$ is theoretically predicted and $\alpha$ extracted from Fig. \ref{fig2}(b)]. This description unifies both flow properties and arrest. Quantitatively, we find that the scaling function $g$ is well-approximated by $g(x)=
 a + \frac{bc x^{\gamma+\beta} }{bx^{\beta}+ c x ^{\gamma}}$ with $a=0.15$, $b=2.0$, $c = 29$ and $\gamma = 0.73$ [Black-line in Fig. \ref{fig4}(b)].


\noindent{\bf Diverging length scale:} We now propose a scaling argument for the exponent $\alpha\approx 0.9$ characterizing the observed finite-size effects in frictionless granular flows. 
We use two facts: (i) in packing of particles, there exists a diverging ``isostatic length-scale'' $\ell^*\sim d/|\delta z|$ (where $\delta z=z-z_c^\infty$ and $z_c^\infty$ is the threshold ``isostatic" coordination where the infinite system rigidifies \cite{maxwell1864})  that characterises how pinning a boundary affects linear properties of the material. Specifically, $\ell^*$ controls both the elasticity of over-constrained materials \cite{wyart2005geometric}, and the scale on which pinning boundaries rigidify a floppy, under-constrained system \cite{C2SM25878A}. We posit that $\ell^*$ is also the length scale on which a granular flow is affected by the presence of a boundary, i.e. $h(\theta_c)\sim \ell^*\sim d/|\delta z|$. (ii) In flowing systems of hard frictionless particles, increasing the macroscopic friction breaks more contact and reduces the coordination. The microscopic theory of \cite{degiuli2015} predicts $|\delta z| \sim \Delta\mu^{0.83}$ both for over-damped and inertial flows, in good agreements with numerics. Putting these two results together, we obtain that jamming occurs when the added constraints from the boundary balance the degrees of freedom of the bulk, leading to $h(\theta_c)\sim d \Delta\mu^{-\alpha}$ with $\alpha=0.83$, in close agreement with our observation $\alpha=0.9\pm 0.1$. Note that another popular length scale that diverges near jamming,  $\ell_c\sim d/\sqrt{|\delta z|}$ \cite{silbert2005vibrations} that characterizes the response to a point perturbation \cite{C2SM25878A}, is ruled out as it would lead to $\alpha=0.41$ incompatible with our data.

\noindent{\bf Discussion:} In this paper, we have shown experimentally that a finite-size frictionless granular layer is devoid of hysteresis: a granular layer of thickness $h/d$ stops or starts flowing at unique critical stability angle $\theta_c(h/d)$. Nonetheless, thinner layers are more stable indicating that finite-size effects are still in play even without solid friction, as previously observed with foams \cite{lespiat2011}. These findings highlight that no hysteretic behaviors emerge from finite-size effects; hysteresis and finite-size effects are thus independent phenomenologies. The absence of hysteresis for a frictionless granular layer, observed both in infinite \cite{perrin2019} and finite size systems, confirms that hysteresis of the avalanche angle is a feature entirely rooted to the presence of inter-particle friction \cite{degiuli2015,perrin2019}. 

The critical stability angle follows $h/d \sim (\tan \theta_c - \tan \theta_c^\infty)^{-\alpha}$ with $\alpha \approx 0.9 \pm0.1$ [Eq. \ref{stabAng} and Fig. \ref{fig2}(b)].  Interestingly, the same empirical law fits rather well previous results, also obtained on an inclined plane but with inertial frictional particles \cite{Pouliquen1999,wyart2009}. Clearly, our results reject mechanisms based on dilatancy effects as frictionless granular layers are not dilatant \cite{PeyneauRoux2008}. The unicity of the critical curve separating flow and arrest ($h_{\rm start}=h_{\rm stop}$) also challenges mechanisms based on flow-induced mechanical noise  \cite{pouliquen2001slow,lemaitre2002origin,pouliquen2009non} or granular fluidity \cite{kamrin2015}, since dynamic noise is absent when starting from a static configuration.   Our experimental work thus called for a new coherent explanation. Here we have argued that non-local effects are a necessary consequences of the finite-size effects near a critical point, justifying the remarkable collapse of flow curves for different thicknesses on a single master curve. The extracted diverging length scale is consistent with the isostatic length scale following $\ell^*\sim d \Delta\mu^{-0.83}$, which is predicted to hold also  for inertial frictionless systems \cite{degiuli2015}.

Future works should investigate whether such generic framework, here validated on inclined planes, also applies to other configurations with boundaries, such as plane shear or heap flow confined between walls \cite{jop2005crucial}.  Another key question is the generalization of these results to the frictional case. In \cite{degiuli2017friction} it was argued that the role of excess coordination in frictionless particles is then replaced by the fraction of sliding contact for frictional ones. The latter appears to be proportional to $\Delta \mu$. Following the argument of the length scale above, we would then obtain $\alpha=1$ which is consistent with observations  \cite{Pouliquen1999,wyart2009}.\vspace{-0.4cm}

\begin{acknowledgments}
\vspace{-0.3cm} We thank O. Pouliquen and S. Mandal for discussions, S. No\"el for building the incline. Support: European Research Council under the European Union Horizon 2020 Research and Innovation programme  (ERC grant agreement No. 647384), ANR \emph{ScienceFriction} (ANR-18-CE30-0024), by the Swiss National Science Foundation for support under Grant No. 200021-165509 and the Simons Foundation Grant (No. 454953 Matthieu Wyart).\end{acknowledgments}

\bibliography{biblio}

\begin{thebibliography}{42}%
\makeatletter
\providecommand \@ifxundefined [1]{%
 \@ifx{#1\undefined}
}%
\providecommand \@ifnum [1]{%
 \ifnum #1\expandafter \@firstoftwo
 \else \expandafter \@secondoftwo
 \fi
}%
\providecommand \@ifx [1]{%
 \ifx #1\expandafter \@firstoftwo
 \else \expandafter \@secondoftwo
 \fi
}%
\providecommand \natexlab [1]{#1}%
\providecommand \enquote  [1]{``#1''}%
\providecommand \bibnamefont  [1]{#1}%
\providecommand \bibfnamefont [1]{#1}%
\providecommand \citenamefont [1]{#1}%
\providecommand \href@noop [0]{\@secondoftwo}%
\providecommand \href [0]{\begingroup \@sanitize@url \@href}%
\providecommand \@href[1]{\@@startlink{#1}\@@href}%
\providecommand \@@href[1]{\endgroup#1\@@endlink}%
\providecommand \@sanitize@url [0]{\catcode `\\12\catcode `\$12\catcode
  `\&12\catcode `\#12\catcode `\^12\catcode `\_12\catcode `\%12\relax}%
\providecommand \@@startlink[1]{}%
\providecommand \@@endlink[0]{}%
\providecommand \url  [0]{\begingroup\@sanitize@url \@url }%
\providecommand \@url [1]{\endgroup\@href {#1}{\urlprefix }}%
\providecommand \urlprefix  [0]{URL }%
\providecommand \Eprint [0]{\href }%
\providecommand \doibase [0]{http://dx.doi.org/}%
\providecommand \selectlanguage [0]{\@gobble}%
\providecommand \bibinfo  [0]{\@secondoftwo}%
\providecommand \bibfield  [0]{\@secondoftwo}%
\providecommand \translation [1]{[#1]}%
\providecommand \BibitemOpen [0]{}%
\providecommand \bibitemStop [0]{}%
\providecommand \bibitemNoStop [0]{.\EOS\space}%
\providecommand \EOS [0]{\spacefactor3000\relax}%
\providecommand \BibitemShut  [1]{\csname bibitem#1\endcsname}%
\let\auto@bib@innerbib\@empty
\bibitem [{\citenamefont {O'Hern}\ \emph {et~al.}(2003)\citenamefont {O'Hern},
  \citenamefont {Silbert}, \citenamefont {Liu},\ and\ \citenamefont
  {Nagel}}]{PhysRevE68011306}%
  \BibitemOpen
  \bibfield  {author} {\bibinfo {author} {\bibfnamefont {C.~S.}\ \bibnamefont
  {O'Hern}}, \bibinfo {author} {\bibfnamefont {L.~E.}\ \bibnamefont {Silbert}},
  \bibinfo {author} {\bibfnamefont {A.~J.}\ \bibnamefont {Liu}}, \ and\
  \bibinfo {author} {\bibfnamefont {S.~R.}\ \bibnamefont {Nagel}},\ }\href
  {\doibase 10.1103/PhysRevE.68.011306} {\bibfield  {journal} {\bibinfo
  {journal} {Phys. Rev. E}\ }\textbf {\bibinfo {volume} {68}},\ \bibinfo
  {pages} {011306} (\bibinfo {year} {2003})}\BibitemShut {NoStop}%
\bibitem [{\citenamefont {Olsson}\ and\ \citenamefont
  {Teitel}(2007)}]{PhysRevLett99178001}%
  \BibitemOpen
  \bibfield  {author} {\bibinfo {author} {\bibfnamefont {P.}~\bibnamefont
  {Olsson}}\ and\ \bibinfo {author} {\bibfnamefont {S.}~\bibnamefont
  {Teitel}},\ }\href {\doibase 10.1103/PhysRevLett.99.178001} {\bibfield
  {journal} {\bibinfo  {journal} {Phys. Rev. Lett.}\ }\textbf {\bibinfo
  {volume} {99}},\ \bibinfo {pages} {178001} (\bibinfo {year}
  {2007})}\BibitemShut {NoStop}%
\bibitem [{\citenamefont {Forterre}\ and\ \citenamefont
  {Pouliquen}(2008)}]{forterre2008}%
  \BibitemOpen
  \bibfield  {author} {\bibinfo {author} {\bibfnamefont {Y.}~\bibnamefont
  {Forterre}}\ and\ \bibinfo {author} {\bibfnamefont {O.}~\bibnamefont
  {Pouliquen}},\ }\href@noop {} {\bibfield  {journal} {\bibinfo  {journal}
  {Annu. Rev. Fluid Mech.}\ }\textbf {\bibinfo {volume} {40}},\ \bibinfo
  {pages} {1} (\bibinfo {year} {2008})}\BibitemShut {NoStop}%
\bibitem [{\citenamefont {Lespiat}\ \emph {et~al.}(2011)\citenamefont
  {Lespiat}, \citenamefont {Cohen-Addad},\ and\ \citenamefont
  {H{\"o}hler}}]{lespiat2011}%
  \BibitemOpen
  \bibfield  {author} {\bibinfo {author} {\bibfnamefont {R.}~\bibnamefont
  {Lespiat}}, \bibinfo {author} {\bibfnamefont {S.}~\bibnamefont
  {Cohen-Addad}}, \ and\ \bibinfo {author} {\bibfnamefont {R.}~\bibnamefont
  {H{\"o}hler}},\ }\href@noop {} {\bibfield  {journal} {\bibinfo  {journal}
  {Physical review letters}\ }\textbf {\bibinfo {volume} {106}},\ \bibinfo
  {pages} {148302} (\bibinfo {year} {2011})}\BibitemShut {NoStop}%
\bibitem [{\citenamefont {Goyon}\ \emph {et~al.}(2010)\citenamefont {Goyon},
  \citenamefont {Colin},\ and\ \citenamefont {Bocquet}}]{goyon2010}%
  \BibitemOpen
  \bibfield  {author} {\bibinfo {author} {\bibfnamefont {J.}~\bibnamefont
  {Goyon}}, \bibinfo {author} {\bibfnamefont {A.}~\bibnamefont {Colin}}, \ and\
  \bibinfo {author} {\bibfnamefont {L.}~\bibnamefont {Bocquet}},\ }\href@noop
  {} {\bibfield  {journal} {\bibinfo  {journal} {Soft Matter}\ }\textbf
  {\bibinfo {volume} {6}},\ \bibinfo {pages} {2668} (\bibinfo {year}
  {2010})}\BibitemShut {NoStop}%
\bibitem [{\citenamefont {Leeman}\ \emph {et~al.}(2016)\citenamefont {Leeman},
  \citenamefont {Saffer}, \citenamefont {Scuderi},\ and\ \citenamefont
  {Marone}}]{leeman2016}%
  \BibitemOpen
  \bibfield  {author} {\bibinfo {author} {\bibfnamefont {J.}~\bibnamefont
  {Leeman}}, \bibinfo {author} {\bibfnamefont {D.}~\bibnamefont {Saffer}},
  \bibinfo {author} {\bibfnamefont {M.}~\bibnamefont {Scuderi}}, \ and\
  \bibinfo {author} {\bibfnamefont {C.}~\bibnamefont {Marone}},\ }\href@noop {}
  {\bibfield  {journal} {\bibinfo  {journal} {Nature communications}\ }\textbf
  {\bibinfo {volume} {7}},\ \bibinfo {pages} {1} (\bibinfo {year}
  {2016})}\BibitemShut {NoStop}%
\bibitem [{\citenamefont {Lucas}\ \emph {et~al.}(2014)\citenamefont {Lucas},
  \citenamefont {Mangeney},\ and\ \citenamefont {Ampuero}}]{lucas2014}%
  \BibitemOpen
  \bibfield  {author} {\bibinfo {author} {\bibfnamefont {A.}~\bibnamefont
  {Lucas}}, \bibinfo {author} {\bibfnamefont {A.}~\bibnamefont {Mangeney}}, \
  and\ \bibinfo {author} {\bibfnamefont {J.~P.}\ \bibnamefont {Ampuero}},\
  }\href@noop {} {\bibfield  {journal} {\bibinfo  {journal} {Nature
  communications}\ }\textbf {\bibinfo {volume} {5}},\ \bibinfo {pages} {1}
  (\bibinfo {year} {2014})}\BibitemShut {NoStop}%
\bibitem [{\citenamefont {Ferdowsi}\ \emph {et~al.}(2018)\citenamefont
  {Ferdowsi}, \citenamefont {Ortiz},\ and\ \citenamefont
  {Jerolmack}}]{ferdowsi2018}%
  \BibitemOpen
  \bibfield  {author} {\bibinfo {author} {\bibfnamefont {B.}~\bibnamefont
  {Ferdowsi}}, \bibinfo {author} {\bibfnamefont {C.~P.}\ \bibnamefont {Ortiz}},
  \ and\ \bibinfo {author} {\bibfnamefont {D.~J.}\ \bibnamefont {Jerolmack}},\
  }\href@noop {} {\bibfield  {journal} {\bibinfo  {journal} {Proceedings of the
  National Academy of Sciences}\ }\textbf {\bibinfo {volume} {115}},\ \bibinfo
  {pages} {4827} (\bibinfo {year} {2018})}\BibitemShut {NoStop}%
\bibitem [{\citenamefont {Pouliquen}(1999)}]{Pouliquen1999}%
  \BibitemOpen
  \bibfield  {author} {\bibinfo {author} {\bibfnamefont {O.}~\bibnamefont
  {Pouliquen}},\ }\href@noop {} {\bibfield  {journal} {\bibinfo  {journal}
  {Physics of fluids}\ }\textbf {\bibinfo {volume} {11}},\ \bibinfo {pages}
  {542} (\bibinfo {year} {1999})}\BibitemShut {NoStop}%
\bibitem [{\citenamefont {Daerr}\ and\ \citenamefont
  {Douady}(1999)}]{daerr1999two}%
  \BibitemOpen
  \bibfield  {author} {\bibinfo {author} {\bibfnamefont {A.}~\bibnamefont
  {Daerr}}\ and\ \bibinfo {author} {\bibfnamefont {S.}~\bibnamefont {Douady}},\
  }\href@noop {} {\bibfield  {journal} {\bibinfo  {journal} {Nature}\ }\textbf
  {\bibinfo {volume} {399}},\ \bibinfo {pages} {241} (\bibinfo {year}
  {1999})}\BibitemShut {NoStop}%
\bibitem [{\citenamefont {Pouliquen}\ and\ \citenamefont
  {Renaut}(1996)}]{Pouliquen1996}%
  \BibitemOpen
  \bibfield  {author} {\bibinfo {author} {\bibfnamefont {O.}~\bibnamefont
  {Pouliquen}}\ and\ \bibinfo {author} {\bibfnamefont {N.}~\bibnamefont
  {Renaut}},\ }\href@noop {} {\bibfield  {journal} {\bibinfo  {journal}
  {Journal de Physique II}\ }\textbf {\bibinfo {volume} {6}},\ \bibinfo {pages}
  {923} (\bibinfo {year} {1996})}\BibitemShut {NoStop}%
\bibitem [{\citenamefont {Erta{\c{s}}}\ and\ \citenamefont
  {Halsey}(2002)}]{ertacs2002}%
  \BibitemOpen
  \bibfield  {author} {\bibinfo {author} {\bibfnamefont {D.}~\bibnamefont
  {Erta{\c{s}}}}\ and\ \bibinfo {author} {\bibfnamefont {T.~C.}\ \bibnamefont
  {Halsey}},\ }\href@noop {} {\bibfield  {journal} {\bibinfo  {journal} {EPL
  (Europhysics Letters)}\ }\textbf {\bibinfo {volume} {60}},\ \bibinfo {pages}
  {931} (\bibinfo {year} {2002})}\BibitemShut {NoStop}%
\bibitem [{\citenamefont {Pouliquen}(2004)}]{pouliquen2004velocity}%
  \BibitemOpen
  \bibfield  {author} {\bibinfo {author} {\bibfnamefont {O.}~\bibnamefont
  {Pouliquen}},\ }\href@noop {} {\bibfield  {journal} {\bibinfo  {journal}
  {Physical review letters}\ }\textbf {\bibinfo {volume} {93}},\ \bibinfo
  {pages} {248001} (\bibinfo {year} {2004})}\BibitemShut {NoStop}%
\bibitem [{\citenamefont {Mills}\ \emph {et~al.}(2008)\citenamefont {Mills},
  \citenamefont {Rognon},\ and\ \citenamefont {Chevoir}}]{mills2008}%
  \BibitemOpen
  \bibfield  {author} {\bibinfo {author} {\bibfnamefont {P.}~\bibnamefont
  {Mills}}, \bibinfo {author} {\bibfnamefont {P.}~\bibnamefont {Rognon}}, \
  and\ \bibinfo {author} {\bibfnamefont {F.}~\bibnamefont {Chevoir}},\
  }\href@noop {} {\bibfield  {journal} {\bibinfo  {journal} {EPL (Europhysics
  Letters)}\ }\textbf {\bibinfo {volume} {81}},\ \bibinfo {pages} {64005}
  (\bibinfo {year} {2008})}\BibitemShut {NoStop}%
\bibitem [{\citenamefont {Gueudr{\'e}}\ \emph {et~al.}(2017)\citenamefont
  {Gueudr{\'e}}, \citenamefont {Lin}, \citenamefont {Rosso},\ and\
  \citenamefont {Wyart}}]{gueudre2017scaling}%
  \BibitemOpen
  \bibfield  {author} {\bibinfo {author} {\bibfnamefont {T.}~\bibnamefont
  {Gueudr{\'e}}}, \bibinfo {author} {\bibfnamefont {J.}~\bibnamefont {Lin}},
  \bibinfo {author} {\bibfnamefont {A.}~\bibnamefont {Rosso}}, \ and\ \bibinfo
  {author} {\bibfnamefont {M.}~\bibnamefont {Wyart}},\ }\href@noop {}
  {\bibfield  {journal} {\bibinfo  {journal} {Soft Matter}\ }\textbf {\bibinfo
  {volume} {13}},\ \bibinfo {pages} {3794} (\bibinfo {year}
  {2017})}\BibitemShut {NoStop}%
\bibitem [{\citenamefont {Pouliquen}\ \emph {et~al.}(2001)\citenamefont
  {Pouliquen}, \citenamefont {Forterre},\ and\ \citenamefont
  {Le~Dizes}}]{pouliquen2001slow}%
  \BibitemOpen
  \bibfield  {author} {\bibinfo {author} {\bibfnamefont {O.}~\bibnamefont
  {Pouliquen}}, \bibinfo {author} {\bibfnamefont {Y.}~\bibnamefont {Forterre}},
  \ and\ \bibinfo {author} {\bibfnamefont {S.}~\bibnamefont {Le~Dizes}},\
  }\href@noop {} {\bibfield  {journal} {\bibinfo  {journal} {Advances in
  complex systems}\ }\textbf {\bibinfo {volume} {4}},\ \bibinfo {pages} {441}
  (\bibinfo {year} {2001})}\BibitemShut {NoStop}%
\bibitem [{\citenamefont {Lemaitre}(2002)}]{lemaitre2002origin}%
  \BibitemOpen
  \bibfield  {author} {\bibinfo {author} {\bibfnamefont {A.}~\bibnamefont
  {Lemaitre}},\ }\href@noop {} {\bibfield  {journal} {\bibinfo  {journal}
  {Physical review letters}\ }\textbf {\bibinfo {volume} {89}},\ \bibinfo
  {pages} {064303} (\bibinfo {year} {2002})}\BibitemShut {NoStop}%
\bibitem [{\citenamefont {Aranson}\ and\ \citenamefont
  {Tsimring}(2002)}]{aranson2002continuum}%
  \BibitemOpen
  \bibfield  {author} {\bibinfo {author} {\bibfnamefont {I.~S.}\ \bibnamefont
  {Aranson}}\ and\ \bibinfo {author} {\bibfnamefont {L.~S.}\ \bibnamefont
  {Tsimring}},\ }\href@noop {} {\bibfield  {journal} {\bibinfo  {journal}
  {Physical Review E}\ }\textbf {\bibinfo {volume} {65}},\ \bibinfo {pages}
  {061303} (\bibinfo {year} {2002})}\BibitemShut {NoStop}%
\bibitem [{\citenamefont {Pouliquen}\ and\ \citenamefont
  {Forterre}(2009)}]{pouliquen2009non}%
  \BibitemOpen
  \bibfield  {author} {\bibinfo {author} {\bibfnamefont {O.}~\bibnamefont
  {Pouliquen}}\ and\ \bibinfo {author} {\bibfnamefont {Y.}~\bibnamefont
  {Forterre}},\ }\href@noop {} {\bibfield  {journal} {\bibinfo  {journal}
  {Philosophical Transactions of the Royal Society A: Mathematical, Physical
  and Engineering Sciences}\ }\textbf {\bibinfo {volume} {367}},\ \bibinfo
  {pages} {5091} (\bibinfo {year} {2009})}\BibitemShut {NoStop}%
\bibitem [{\citenamefont {Kamrin}\ and\ \citenamefont
  {Henann}(2015)}]{kamrin2015}%
  \BibitemOpen
  \bibfield  {author} {\bibinfo {author} {\bibfnamefont {K.}~\bibnamefont
  {Kamrin}}\ and\ \bibinfo {author} {\bibfnamefont {D.~L.}\ \bibnamefont
  {Henann}},\ }\href@noop {} {\bibfield  {journal} {\bibinfo  {journal} {Soft
  matter}\ }\textbf {\bibinfo {volume} {11}},\ \bibinfo {pages} {179} (\bibinfo
  {year} {2015})}\BibitemShut {NoStop}%
\bibitem [{\citenamefont {Wyart}(2009)}]{wyart2009}%
  \BibitemOpen
  \bibfield  {author} {\bibinfo {author} {\bibfnamefont {M.}~\bibnamefont
  {Wyart}},\ }\href@noop {} {\bibfield  {journal} {\bibinfo  {journal} {EPL
  (Europhysics Letters)}\ }\textbf {\bibinfo {volume} {85}},\ \bibinfo {pages}
  {24003} (\bibinfo {year} {2009})}\BibitemShut {NoStop}%
\bibitem [{\citenamefont {Edwards}\ \emph {et~al.}(2019)\citenamefont
  {Edwards}, \citenamefont {Russell}, \citenamefont {Johnson},\ and\
  \citenamefont {Gray}}]{edwards2019}%
  \BibitemOpen
  \bibfield  {author} {\bibinfo {author} {\bibfnamefont {A.}~\bibnamefont
  {Edwards}}, \bibinfo {author} {\bibfnamefont {A.}~\bibnamefont {Russell}},
  \bibinfo {author} {\bibfnamefont {C.}~\bibnamefont {Johnson}}, \ and\
  \bibinfo {author} {\bibfnamefont {J.}~\bibnamefont {Gray}},\ }\href@noop {}
  {\bibfield  {journal} {\bibinfo  {journal} {Journal of Fluid Mechanics}\
  }\textbf {\bibinfo {volume} {875}},\ \bibinfo {pages} {1058} (\bibinfo {year}
  {2019})}\BibitemShut {NoStop}%
\bibitem [{\citenamefont {Clavaud}\ \emph {et~al.}(2017)\citenamefont
  {Clavaud}, \citenamefont {B{\'e}rut}, \citenamefont {Metzger},\ and\
  \citenamefont {Forterre}}]{Clavaud2017}%
  \BibitemOpen
  \bibfield  {author} {\bibinfo {author} {\bibfnamefont {C.}~\bibnamefont
  {Clavaud}}, \bibinfo {author} {\bibfnamefont {A.}~\bibnamefont {B{\'e}rut}},
  \bibinfo {author} {\bibfnamefont {B.}~\bibnamefont {Metzger}}, \ and\
  \bibinfo {author} {\bibfnamefont {Y.}~\bibnamefont {Forterre}},\ }\href
  {\doibase 10.1073/pnas.1703926114} {\bibfield  {journal} {\bibinfo  {journal}
  {Proceedings of the National Academy of Sciences}\ }\textbf {\bibinfo
  {volume} {114}},\ \bibinfo {pages} {5147} (\bibinfo {year}
  {2017})}\BibitemShut {NoStop}%
\bibitem [{\citenamefont {Perrin}\ \emph {et~al.}(2019)\citenamefont {Perrin},
  \citenamefont {Clavaud}, \citenamefont {Wyart}, \citenamefont {Metzger},\
  and\ \citenamefont {Forterre}}]{perrin2019}%
  \BibitemOpen
  \bibfield  {author} {\bibinfo {author} {\bibfnamefont {H.}~\bibnamefont
  {Perrin}}, \bibinfo {author} {\bibfnamefont {C.}~\bibnamefont {Clavaud}},
  \bibinfo {author} {\bibfnamefont {M.}~\bibnamefont {Wyart}}, \bibinfo
  {author} {\bibfnamefont {B.}~\bibnamefont {Metzger}}, \ and\ \bibinfo
  {author} {\bibfnamefont {Y.}~\bibnamefont {Forterre}},\ }\href@noop {}
  {\bibfield  {journal} {\bibinfo  {journal} {Physical Review X}\ }\textbf
  {\bibinfo {volume} {9}},\ \bibinfo {pages} {031027} (\bibinfo {year}
  {2019})}\BibitemShut {NoStop}%
\bibitem [{\citenamefont {Peyneau}\ and\ \citenamefont
  {Roux}(2008)}]{PeyneauRoux2008}%
  \BibitemOpen
  \bibfield  {author} {\bibinfo {author} {\bibfnamefont {P.-E.}\ \bibnamefont
  {Peyneau}}\ and\ \bibinfo {author} {\bibfnamefont {J.-N.}\ \bibnamefont
  {Roux}},\ }\href {\doibase 10.1103/PhysRevE.78.011307} {\bibfield  {journal}
  {\bibinfo  {journal} {Phys. Rev. E}\ }\textbf {\bibinfo {volume} {78}},\
  \bibinfo {pages} {011307} (\bibinfo {year} {2008})}\BibitemShut {NoStop}%
\bibitem [{\citenamefont {DeGiuli}\ \emph {et~al.}(2015)\citenamefont
  {DeGiuli}, \citenamefont {D{\"u}ring}, \citenamefont {Lerner},\ and\
  \citenamefont {Wyart}}]{degiuli2015}%
  \BibitemOpen
  \bibfield  {author} {\bibinfo {author} {\bibfnamefont {E.}~\bibnamefont
  {DeGiuli}}, \bibinfo {author} {\bibfnamefont {G.}~\bibnamefont {D{\"u}ring}},
  \bibinfo {author} {\bibfnamefont {E.}~\bibnamefont {Lerner}}, \ and\ \bibinfo
  {author} {\bibfnamefont {M.}~\bibnamefont {Wyart}},\ }\href@noop {}
  {\bibfield  {journal} {\bibinfo  {journal} {Physical Review E}\ }\textbf
  {\bibinfo {volume} {91}},\ \bibinfo {pages} {062206} (\bibinfo {year}
  {2015})}\BibitemShut {NoStop}%
\bibitem [{\citenamefont {Combe}\ and\ \citenamefont {Roux}(2000)}]{combe2000}%
  \BibitemOpen
  \bibfield  {author} {\bibinfo {author} {\bibfnamefont {G.}~\bibnamefont
  {Combe}}\ and\ \bibinfo {author} {\bibfnamefont {J.-N.}\ \bibnamefont
  {Roux}},\ }\href@noop {} {\bibfield  {journal} {\bibinfo  {journal} {Physical
  Review Letters}\ }\textbf {\bibinfo {volume} {85}},\ \bibinfo {pages} {3628}
  (\bibinfo {year} {2000})}\BibitemShut {NoStop}%
\bibitem [{\citenamefont {Lerner}\ \emph {et~al.}(2012)\citenamefont {Lerner},
  \citenamefont {D{\"u}ring},\ and\ \citenamefont {Wyart}}]{lerner2012}%
  \BibitemOpen
  \bibfield  {author} {\bibinfo {author} {\bibfnamefont {E.}~\bibnamefont
  {Lerner}}, \bibinfo {author} {\bibfnamefont {G.}~\bibnamefont {D{\"u}ring}},
  \ and\ \bibinfo {author} {\bibfnamefont {M.}~\bibnamefont {Wyart}},\
  }\href@noop {} {\bibfield  {journal} {\bibinfo  {journal} {Proceedings of the
  National Academy of Sciences}\ }\textbf {\bibinfo {volume} {109}},\ \bibinfo
  {pages} {4798} (\bibinfo {year} {2012})}\BibitemShut {NoStop}%
\bibitem [{\citenamefont {Gallier}\ \emph {et~al.}(2014)\citenamefont
  {Gallier}, \citenamefont {Lemaire}, \citenamefont {Peters},\ and\
  \citenamefont {Lobry}}]{gallier2014}%
  \BibitemOpen
  \bibfield  {author} {\bibinfo {author} {\bibfnamefont {S.}~\bibnamefont
  {Gallier}}, \bibinfo {author} {\bibfnamefont {E.}~\bibnamefont {Lemaire}},
  \bibinfo {author} {\bibfnamefont {F.}~\bibnamefont {Peters}}, \ and\ \bibinfo
  {author} {\bibfnamefont {L.}~\bibnamefont {Lobry}},\ }\href@noop {}
  {\bibfield  {journal} {\bibinfo  {journal} {Journal of Fluid Mechanics}\
  }\textbf {\bibinfo {volume} {757}},\ \bibinfo {pages} {514} (\bibinfo {year}
  {2014})}\BibitemShut {NoStop}%
\bibitem [{\citenamefont {Mari}\ \emph {et~al.}(2014)\citenamefont {Mari},
  \citenamefont {Seto}, \citenamefont {Morris},\ and\ \citenamefont
  {Denn}}]{mari2014}%
  \BibitemOpen
  \bibfield  {author} {\bibinfo {author} {\bibfnamefont {R.}~\bibnamefont
  {Mari}}, \bibinfo {author} {\bibfnamefont {R.}~\bibnamefont {Seto}}, \bibinfo
  {author} {\bibfnamefont {J.~F.}\ \bibnamefont {Morris}}, \ and\ \bibinfo
  {author} {\bibfnamefont {M.~M.}\ \bibnamefont {Denn}},\ }\href@noop {}
  {\bibfield  {journal} {\bibinfo  {journal} {Journal of Rheology}\ }\textbf
  {\bibinfo {volume} {58}},\ \bibinfo {pages} {1693} (\bibinfo {year}
  {2014})}\BibitemShut {NoStop}%
\bibitem [{\citenamefont {Wyart}\ and\ \citenamefont
  {Cates}(2014)}]{wyart2014discontinuous}%
  \BibitemOpen
  \bibfield  {author} {\bibinfo {author} {\bibfnamefont {M.}~\bibnamefont
  {Wyart}}\ and\ \bibinfo {author} {\bibfnamefont {M.}~\bibnamefont {Cates}},\
  }\href@noop {} {\bibfield  {journal} {\bibinfo  {journal} {Physical review
  letters}\ }\textbf {\bibinfo {volume} {112}},\ \bibinfo {pages} {098302}
  (\bibinfo {year} {2014})}\BibitemShut {NoStop}%
\bibitem [{\citenamefont {Cardy}(2012)}]{cardy2012finite}%
  \BibitemOpen
  \bibfield  {author} {\bibinfo {author} {\bibfnamefont {J.}~\bibnamefont
  {Cardy}},\ }\href@noop {} {\emph {\bibinfo {title} {Finite-size scaling}}}\
  (\bibinfo  {publisher} {Elsevier},\ \bibinfo {year} {2012})\BibitemShut
  {NoStop}%
\bibitem [{\citenamefont {Wyart}\ \emph {et~al.}(2005)\citenamefont {Wyart},
  \citenamefont {Nagel},\ and\ \citenamefont {Witten}}]{wyart2005geometric}%
  \BibitemOpen
  \bibfield  {author} {\bibinfo {author} {\bibfnamefont {M.}~\bibnamefont
  {Wyart}}, \bibinfo {author} {\bibfnamefont {S.~R.}\ \bibnamefont {Nagel}}, \
  and\ \bibinfo {author} {\bibfnamefont {T.~A.}\ \bibnamefont {Witten}},\
  }\href@noop {} {\bibfield  {journal} {\bibinfo  {journal} {EPL (Europhysics
  Letters)}\ }\textbf {\bibinfo {volume} {72}},\ \bibinfo {pages} {486}
  (\bibinfo {year} {2005})}\BibitemShut {NoStop}%
\bibitem [{\citenamefont {Israelachvili}(2011)}]{Israelachvili2011Book}%
  \BibitemOpen
  \bibfield  {author} {\bibinfo {author} {\bibfnamefont {J.~N.}\ \bibnamefont
  {Israelachvili}},\ }\href@noop {} {\emph {\bibinfo {title} {Intermolecular
  and surface forces}}}\ (\bibinfo  {publisher} {Academic press},\ \bibinfo
  {year} {2011})\BibitemShut {NoStop}%
\bibitem [{\citenamefont {Cassar}\ \emph {et~al.}(2005)\citenamefont {Cassar},
  \citenamefont {Nicolas},\ and\ \citenamefont
  {Pouliquen}}]{cassar2005submarine}%
  \BibitemOpen
  \bibfield  {author} {\bibinfo {author} {\bibfnamefont {C.}~\bibnamefont
  {Cassar}}, \bibinfo {author} {\bibfnamefont {M.}~\bibnamefont {Nicolas}}, \
  and\ \bibinfo {author} {\bibfnamefont {O.}~\bibnamefont {Pouliquen}},\
  }\href@noop {} {\bibfield  {journal} {\bibinfo  {journal} {Physics of
  fluids}\ }\textbf {\bibinfo {volume} {17}},\ \bibinfo {pages} {103301}
  (\bibinfo {year} {2005})}\BibitemShut {NoStop}%
\bibitem [{\citenamefont {Erpelding}\ \emph {et~al.}(2008)\citenamefont
  {Erpelding}, \citenamefont {Amon},\ and\ \citenamefont
  {Crassous}}]{CrassousCorrPRE2008}%
  \BibitemOpen
  \bibfield  {author} {\bibinfo {author} {\bibfnamefont {M.}~\bibnamefont
  {Erpelding}}, \bibinfo {author} {\bibfnamefont {A.}~\bibnamefont {Amon}}, \
  and\ \bibinfo {author} {\bibfnamefont {J.}~\bibnamefont {Crassous}},\
  }\href@noop {} {\bibfield  {journal} {\bibinfo  {journal} {Phys. Rev. E}\
  }\textbf {\bibinfo {volume} {78}},\ \bibinfo {pages} {046104} (\bibinfo
  {year} {2008})}\BibitemShut {NoStop}%
\bibitem [{Note1()}]{Note1}%
  \BibitemOpen
  \bibinfo {note} {C is the spatial average of the correlation matrix $M =
  \protect \frac {\left <I(0) I(\Delta t)\right > - \left <I(0)\right >\left
  <I(\Delta t)\right >} {\protect \sqrt {\left <I(0)^2\right > -\left
  <I(0)\right >^2 } \protect \sqrt {\left <I(\Delta t)^2\right > -\left
  <I(\Delta t)\right >^2 } } $, where $I(0)$ and $I(\Delta t)$ are two images
  separated by a time interval $\Delta t = 0.1\protect \tmspace +\thickmuskip
  {.2777em}\protect \rm {s}$ and $\left <.\right >$ denotes the average
  correlation over boxes of $4\times 4$ pixels, see \cite {CrassousCorrPRE2008}
  for details}\BibitemShut {NoStop}%
\bibitem [{\citenamefont {Maxwell}(1864)}]{maxwell1864}%
  \BibitemOpen
  \bibfield  {author} {\bibinfo {author} {\bibfnamefont {J.~C.}\ \bibnamefont
  {Maxwell}},\ }\href@noop {} {\bibfield  {journal} {\bibinfo  {journal} {The
  London, Edinburgh, and Dublin Philosophical Magazine and Journal of Science}\
  }\textbf {\bibinfo {volume} {27}},\ \bibinfo {pages} {294} (\bibinfo {year}
  {1864})}\BibitemShut {NoStop}%
\bibitem [{\citenamefont {D{\"u}ring}\ \emph {et~al.}(2013)\citenamefont
  {D{\"u}ring}, \citenamefont {Lerner},\ and\ \citenamefont
  {Wyart}}]{C2SM25878A}%
  \BibitemOpen
  \bibfield  {author} {\bibinfo {author} {\bibfnamefont {G.}~\bibnamefont
  {D{\"u}ring}}, \bibinfo {author} {\bibfnamefont {E.}~\bibnamefont {Lerner}},
  \ and\ \bibinfo {author} {\bibfnamefont {M.}~\bibnamefont {Wyart}},\
  }\href@noop {} {\bibfield  {journal} {\bibinfo  {journal} {Soft Matter}\
  }\textbf {\bibinfo {volume} {9}},\ \bibinfo {pages} {146} (\bibinfo {year}
  {2013})}\BibitemShut {NoStop}%
\bibitem [{\citenamefont {Silbert}\ \emph {et~al.}(2005)\citenamefont
  {Silbert}, \citenamefont {Liu},\ and\ \citenamefont
  {Nagel}}]{silbert2005vibrations}%
  \BibitemOpen
  \bibfield  {author} {\bibinfo {author} {\bibfnamefont {L.~E.}\ \bibnamefont
  {Silbert}}, \bibinfo {author} {\bibfnamefont {A.~J.}\ \bibnamefont {Liu}}, \
  and\ \bibinfo {author} {\bibfnamefont {S.~R.}\ \bibnamefont {Nagel}},\
  }\href@noop {} {\bibfield  {journal} {\bibinfo  {journal} {Physical review
  letters}\ }\textbf {\bibinfo {volume} {95}},\ \bibinfo {pages} {098301}
  (\bibinfo {year} {2005})}\BibitemShut {NoStop}%
\bibitem [{\citenamefont {Jop}\ \emph {et~al.}(2005)\citenamefont {Jop},
  \citenamefont {Forterre},\ and\ \citenamefont {Pouliquen}}]{jop2005crucial}%
  \BibitemOpen
  \bibfield  {author} {\bibinfo {author} {\bibfnamefont {P.}~\bibnamefont
  {Jop}}, \bibinfo {author} {\bibfnamefont {Y.}~\bibnamefont {Forterre}}, \
  and\ \bibinfo {author} {\bibfnamefont {O.}~\bibnamefont {Pouliquen}},\
  }\href@noop {} {\bibfield  {journal} {\bibinfo  {journal} {arXiv preprint
  cond-mat/0503425}\ } (\bibinfo {year} {2005})}\BibitemShut {NoStop}%
\bibitem [{\citenamefont {DeGiuli}\ and\ \citenamefont
  {Wyart}(2017)}]{degiuli2017friction}%
  \BibitemOpen
  \bibfield  {author} {\bibinfo {author} {\bibfnamefont {E.}~\bibnamefont
  {DeGiuli}}\ and\ \bibinfo {author} {\bibfnamefont {M.}~\bibnamefont
  {Wyart}},\ }\href@noop {} {\bibfield  {journal} {\bibinfo  {journal}
  {Proceedings of the National Academy of Sciences}\ }\textbf {\bibinfo
  {volume} {114}},\ \bibinfo {pages} {9284} (\bibinfo {year}
  {2017})}\BibitemShut {NoStop}%
\end{thebibliography}%

\end{document}